**Doxorubicin Loaded Magnetic Polymersomes: Theranostic Nanocarriers for MR Imaging and Magneto-Chemotherapy**

Charles Sanson[a], Odile Diou[a], Julie Thévenot[a], Emmanuel Ibarboure[a], Alain Soum[a], Annie Brûlet[b], Sylvain Miraux[c], Eric Thiaudière[c], Sisareuth Tan[d], Alain Brisson[d], Vincent Dupuis[e], Olivier Sandre[a] and Sébastien Lecommandoux[a]

[a] LCPO UMR5629 Univ Bordeaux / CNRS / IPB,
[b] RMSB UMR5536 Univ Bordeaux / CNRS,
[c] LLB UMR12 CNRS / CEA-Saclay,
[d] CBMN UMR5248 Univ Bordeaux / CNRS / IPB
[e] PECSA UMR7195 UPMC Paris 6 / CNRS
olivier.sandre@ipb.fr, lecommandoux@enscbp.fr

Hydrophobically modified magnetic nanoparticles (MNPs) were encapsulated within the membrane of poly(trimethylene carbonate)-*b*-poly(L-glutamic acid) (PTMC-*b*-PGA) block copolymer vesicles using a nanoprecipitation process. This formulation method provides a high loading of MNPs (up to 70 wt %) together with a good control over the sizes of the vesicles (100 – 400 nm). The deformation of the vesicle membrane under an applied magnetic field was evidenced by anisotropic SANS. These hybrid objects display contrast enhancement properties in Magnetic Resonance Imaging, a diagnostic method routinely used for three-dimensional and non-invasive scans of the human body.[1] They can also be guided in a magnetic field gradient. The feasibility of drug release triggered by magnetic induction was evidenced using the anticancer drug doxorubicin (DOX), which is co-encapsulated in the membrane. Magnetic polymersomes are thus proposed as multimodal drug nanocarriers for bio-imaging and magneto-chemotherapy.

In the new field of nanomedicine, a great variety of polymeric drug delivery nanocarriers have shown efficient entrapment and controlled release of drugs *in vitro*. However, evaluating the fate of the carrier after injection *in vivo* is a delicate task: as a consequence, an imaging probe is usually co-encapsulated with the drugs in the polymer nanoparticles. Such multifunctional nanocarriers for cancer diagnostics and treatment open the field of "theranostics", *i.e.* combination of imaging and therapeutic functions in "all-in-one" nanoparticles.[2] Among the different nanoparticles used in biotechnological applications, superparamagnetic nanocrystals made of $\gamma$-$Fe_2O_3$ around 10 nm in size are of particular interest since they are biocompatible and offer contrast enhancement in MRI. In addition, MNPs can also be implemented to induce cancer cells apoptosis. Local heating is obtained by exposure to radio-frequency magnetic fields in the frequency range 100 kHz – 1MHz, due to thermal dissipation by the magnetic moments forced to oscillate at the frequency of the applied magnetic field. Hyperthermia therapy (*i.e.* treating by heat) has been recognized as a promising form of cancer therapy, particularly in synergy with chemo- or radio-therapy.[3] For a solid tumor located not too deep under the skin (such as melanoma), a conventional hyperthermia treatment is applied, the cancer cells being destroyed by the heat shock. We proposed a totally different strategy with a polymer vesicle delivering a drug through the temperature increase induced by radiofrequency magnetic fields not macroscopically but locally, at the scale of the membrane: the MNPs being confined inside the polymer block, the heat dissipates in the vicinity of the polymer membrane, which strongly impacts its permeability. Therefore we called this strategy "magneto-chemotherapy" as depicted on Fig. 1.[4]

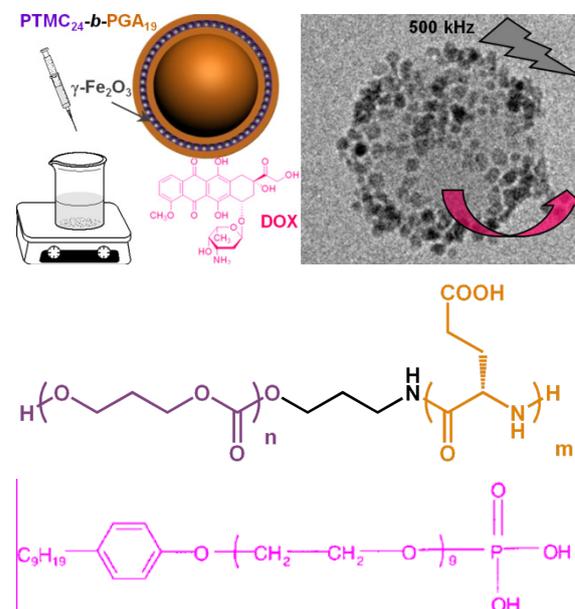

Fig. 1: Dually-loaded vesicles prepared by addition at controlled speed of an aqueous buffer into a mixture in DMSO of PTMC-*b*-PGA (m=19 and n=24), doxorucin and magnetic nanoparticles coated by the BEY surfactant of given formula (bottom sketch). The cryo-TEM image (top right panel) shows a vesicle wrapped by a mantle of MNPs, which excitation by a RF magnetic field heats up the membrane locally and accelerates the DOX release.

In this highlight paper, we describe a convenient way to prepare fully biodegradable polymersomes featuring a dual magnetic and thermo-responsive membrane. The hydrophobic core of membranes is made of the PTMC semi-crystalline blocks and the hydrophilic leaflets arise from the PGA polypeptide blocks of the copolymer. One pot nanoprecipitation (Fig. 1) leads to dual-loaded vesicles with iron oxide MNPs of appropriate size and coating (BEY) embedded within the membranes together with the antitumor drug doxorubicin. The amphiphilic diblock copolymer was synthesized by ring-opening polymerization of $\gamma$-benzyl-L-glutamate *N*-carboxyanhydride initiated by amino functionalized PTMC macroinitiator upon a previously published method.[5] $PTMC_{24}$-*b*-$PGA_{19}$ presents a hydrophilic weight fraction of 50 wt % for an overall molar mass $M_n$ = 4900 g/mol with a dispersity index 1.15. On the other hand, magnetic iron oxides MNPs were synthesized in water by alkaline co-precipitation of ferrous and ferric salts followed by coating with a surfactant called Beycostat (bottom sketch on Fig 1). The gyration and hydrodynamic radii of these MNPs were respectively $R_G^{MNP}$=3.05±0.06 nm and $R_H^{MNP}$=4.70±0.07 nm as measured by SANS and DLS.

Nanoprecipitation allowed reaching quantitative loading contents (LC) with controlled final sizes and low polydispersity indexes, as reported on this table for two durations of the buffer addition (PBS pH=7.4), with and without DOX.

| Addition time | MNP LC | DOX LC | $R_H$ (nm) | PDI | $\zeta$ (mV) |
|---|---|---|---|---|---|
| 15 min | 35 % | 0 % | 152 | 0.15 | -39.6 |
|  |  | 12 % | 124 | 0.23 | -39.3 |
| 10 sec | 50 % | 0 % | 56.5 | 0.22 | -40.8 |
|  |  | 9 % | 61 | 0.15 | -42.0 |

The two-dimensional confinement of the MNPs inside the vesicular membranes was evidenced by small angle neutron and light scattering techniques and observed by transmission electron microscopy. The magnetic membranes exhibited a reversible deformation under a permanent magnetic field of moderate intensity (0.1 Tesla) as seen on Fig. 2.

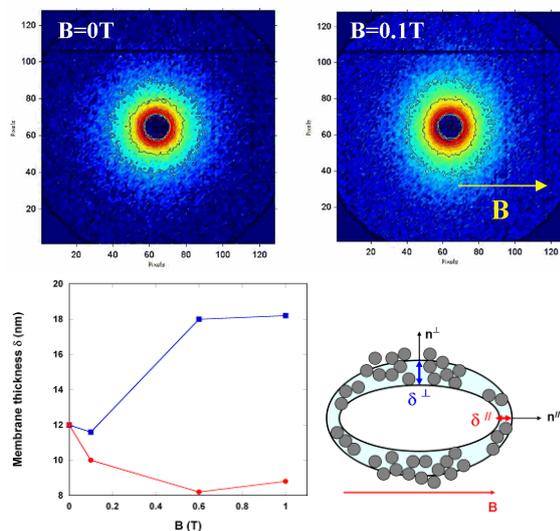

Fig. 2 Top: SANS patterns of magnetic vesicles loaded at 50 wt. % MNPs in $H_2O$ in the $q$ range $3\times10^{-3} – 3\times10^{-2}$ Å$^{-1}$ at zero field and under $B$=0.1 Tesla. Colors correspond to iso-intensity ranges. Patterns were averaged in angular sectors (-30° 30°) along the field direction or (-15° 15°) around the perpendicular direction. Bottom: Kratky-Porod fits $Ln[q^2 I(q)] \sim - q^2 \delta^2 /12$ in the low $q$ regime lead to membrane thicknesses in the two directions $\delta^{//}$ and $\delta^\perp$ and to a scheme of the overall deformation.

With pure $H_2O$ as solvent, the nuclear scattering of the inorganic component of membranes was much greater than that of the polymer leaflets. Thus a variation in intensity in both directions can be ascribed to a reorganization of the MNPs within the membrane. The variation of $\delta^{//}$ and $\delta^\perp$ as a function of the field intensity is interpreted by the stretching of the membrane near the magnetic poles due to an overall deformation of the vesicle into an elongated ellipsoid. The apparent increase of thickness near the equator and concomitant decrease near the magnetic poles ($\delta^\perp$ = 18 nm, $\delta^{//}$ = 8 nm at B=0.6 T) indicates that MNPs move away from the poles where dipolar repulsions are strong and accumulate in the other portions of membrane where dipolar interactions are expected to be attractive

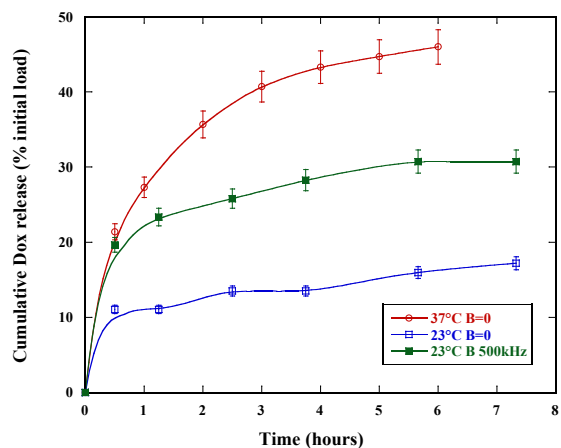

Fig. 3: *In vitro* release assays of doxorubicin at two temperatures (23 or 37°C) and with or without an applied radio-frequency magnetic field B=2.65 mT at 500 kHz.

Finally, the release of the anticancer drug was assayed under several heating conditions. A global heating at 37°C (melting temperature of PTMC[6]) caused a significantly higher release than the one observed at 23°C, as explained by a more permeable membrane. But the application of an oscillating magnetic field at 23°C induced hyperthermia at a local (nanometric) scale only in the vicinity of the membrane that provoked an increase in DOX release (by 2 times), thereby evidencing the validity of the concept of magneto-chemotherapy with those magnetic polymersomes.

**Funding**
Financial support was provided by the European Commission (FP7 CP-IP 213631-2 NANOTHER).